\documentclass[amssymb,amsfonts,aps,prl,twocolumn,showpacs]{revtex4}
\usepackage{graphicx}
\usepackage{dcolumn}
\usepackage{bm}
\newcommand{\beq}{\begin{equation}}  
\newcommand{\eeq}{\end{equation}}  
\newcommand{\beqa}{\begin{eqnarray}}  
\newcommand{\eeqa}{\end{eqnarray}}

\begin{document}

\title{Inelastic electron tunneling spectroscopy of a Mn dimer}

\author{F. Delgado$^1$ and J. Fern\'andez-Rossier$^{1,2}$ }
\affiliation{$^1$ INL-International Iberian Nanotechnology Laboratory,
Av. Mestre Jos\'e Veiga, 4715-310 Braga, Portugal\\
$^2$Departamento de F\'{\i}sica Aplicada,
Universidad de Alicante, San Vicente del Raspeig, 03690 Spain}


\begin{abstract}
A scanning tunneling microscope (STM) can probe the inelastic spin excitations of single magnetic atoms in a surface via spin-flip assisted tunneling. 
A particular and intriguing case is the Mn dimer case. We show here that the existing theories for inelastic transport spectroscopy do not explain the observed spin transitions when both atoms are equally coupled to the STM tip and the substrate. The hyperfine coupling to the nuclear spins is shown to lead to a finite excitation amplitude, but the physical mechanism leading to the large inelastic signal observed is still unknown. We discuss some other alternatives that break the symmetry of the system and allows for larger excitation probabilities.  
\end{abstract}


\maketitle

\section{Model}
The spin of single or few magnetic atoms as well as single magnetic molecules deposited in conducting surfaces can be probed using scanning tunneling
microscopes\cite{Heinrich_Gupta_science_2004,Hirjibehedin_Lutz_Science_2006,Hirjibehedin_Lin_Science_2007,Otte_Ternes_natphys_2008,Loth_Bergmann_natphys_2010,Chen_Fu_prl_2008,Krause_Bautista_Science_2007,Meier_Zhou_Science_2008,Wiesendanger_revmod_2009,Tsukahara_Noto_prl_2009,Fu_Zhang_prl_2009,Brune_Gambardella_sursci_2009,Zhou_Wiebe_natphys_2010}. This can be done with the help of
two complementary techniques: spin polarized STM and spin-flip inelastic electron tunnel spectroscopy (IETS). 
In the first case, control of the spin orientation of either the tip or the substrate permits spin contrast STM imaging\cite{Wiesendanger_revmod_2009}, with the spin dependent 
magneto resistance\cite{Slonczewski_PRB_1989} responsible of the spin contrast. 
 In the case of IETS, electrons tunnel from the  STM tip to the surface (or vice versa), and exchange their spin with the magnetic adatom, producing a spin transition whose energy is provided by the bias voltage.  Whenever a new conduction channel opens with increasing bias voltage, a step in the conductance appears. This permits one to determine the energy of the spin excitation, or how it evolves as a function  of an applied magnetic field.
The observed excitation spectra of Mn, Co and Fe atoms\cite{Heinrich_Gupta_science_2004,Hirjibehedin_Lutz_Science_2006,Hirjibehedin_Lin_Science_2007,Otte_Ternes_natphys_2008} as well as  Fe and Co  
Phthalocyanines molecules\cite{Chen_Fu_prl_2008,Tsukahara_Noto_prl_2009,Fu_Zhang_prl_2009} deposited on an insulating monolayer on top of a metal, have been successfully described using spin Hamiltonians.

Chains of magnetic atoms have also been studied using IETS, where the spin interactions between  the magnetic adatoms in engineered atomic structures were characterized\cite{Hirjibehedin_Lutz_Science_2006,Loth_Bergmann_natphys_2010,Otte_PhD_2008}. The experimentally studied Mn dimer on a Cu$_2$N substrate constitutes a particularly intriguing case. As it was pointed out by us in a previous paper\cite{Delgado_Rossier_prb_2010}, the existing theory of inelastic spin spectroscopy predicts the suppression of the inelastic transitions in the conditions reported in experiments\cite{Hirjibehedin_Lutz_Science_2006,Loth_Bergmann_natphys_2010}. Here we propose the hyperfine coupling between the Mn spins and their nuclear spins as the mechanims responsible of the finite excitation probability.

\section{Model}
The results presented in this Letter are based on a phenomenological (quantized) single ion Hamiltonian
\cite{Heinrich_Gupta_science_2004,Hirjibehedin_Lutz_Science_2006,Hirjibehedin_Lin_Science_2007,Otte_Ternes_natphys_2008,Chen_Fu_prl_2008,Krause_Bautista_Science_2007,Meier_Zhou_Science_2008,Wiesendanger_revmod_2009,Tsukahara_Noto_prl_2009,Fu_Zhang_prl_2009,Brune_Gambardella_sursci_2009,Zhou_Wiebe_natphys_2010}
\begin{equation}
{\cal H}(i)= D \hat{S}_z^{2}(i) + E(\hat{S}_x^{2}(i)-\hat{S}_y^{2}(i))+g\mu_B {\bf \hat{S}}(i).{\bf B}.
\label{Hspini}
\end{equation}
The first term describes the single ion uniaxial magneto-crystalline anisotropy, the second describes the traversal anisotropy and
the third  corresponds to the Zeeman splitting term under an applied magnetic field ${\bf B}$.
Here the $z$-axis corresponds to the easy-axis ($D<0$) or hard-plane ($D>0$) of the adatom.
The value of the spin $S(i)$, and the magnetic anisotropy coefficients $D$ and $E$ change from atom to atom and also depend on the substrate\cite{Hirjibehedin_Lin_Science_2007}. 

The Hamiltonian of a chain of magnetic adatoms can be described by its magnetic anisotropy plus a Heisenberg interaction between them\cite{Hirjibehedin_Lutz_Science_2006,Loth_Bergmann_natphys_2010,Otte_PhD_2008}, 
\beqa
{\cal H}_{Chain}=\sum_i^N {\cal H}(i)+\frac{J}{2}\sum_{\langle i,j\rangle} {\bf S}(i){\bf S}(j),
\label{hchain}
\eeqa
where $J$ is the exchange interaction between tha magnetic atoms and the double sum is over first neighbours. 
The physical system is modeled as two electrodes, tip and substrate, exchange coupled to the chain of magnetic adatoms, ${\cal H}= {\cal H}_{\rm T} + {\cal H}_{\rm S} + {\cal H}_{\rm Chain}+ {\cal V}$.
Here ${\cal H}_{\rm T}+{\cal H}_{\rm S}=\sum_{\lambda\sigma} \epsilon_{\lambda}  
c^{\dagger}_{\lambda\sigma}c_{\lambda\sigma}\; $ describes the electrons in the non-magnetic electrodes, with energy $\epsilon_{\lambda}$, where
$c_{\lambda\sigma}^\dag$ ($c_{\lambda\sigma}$), is the creation (annihilation) operator of a quasiparticle with single particle quantum numbers $(\lambda\sigma)\equiv (k,\eta,\sigma)$, momentum $k$, spin projection $\sigma$, and electrode $\eta={\rm T, S}$. The ${\cal V}$ term introduces the interactions between the three uncoupled systems and enables the transport. It has 
the form\cite{Appelbaum_pr_1967,Rossier_prl_2009,Fransson_nanolett_2009,Lorente_Gauyacq_prl_2009,Delgado_Palacios_prl_2010,Gauyacq_Novaes_prb_2010,Fransson_Eriksson_prb_2010,Zitko_Pruschke_njphys_2010}
\begin{equation}
{\cal V}=
\sum_{\lambda\lambda',\sigma\sigma',i} \left(\frac{T_0}{2} \hat{I}+{\cal T}\sum_{a}\frac{\tau^{(a)}_{\sigma\sigma'}}{2}  \hat{S}_{a}(i)\right)
v_\lambda(i) v_{\lambda'}(i)
c^{\dagger}_{\lambda,\sigma} c_{\lambda'\sigma'}, 
\label{HTUN}
\end{equation}
where $a=x,y,z$.
We use $\tau^{(a)}$ and $\hat{S}_{a}$ for the   Pauli matrices
and  the spin operators, while $\hat{I}$ is the identity matrix.
Neglecting the momentum dependence, which can be safely done for the low bias applied in IETS\cite{Heinrich_Gupta_science_2004,Hirjibehedin_Lutz_Science_2006,Hirjibehedin_Lin_Science_2007,Otte_Ternes_natphys_2008}, one can write $v_\lambda(i) \equiv v_{\eta}(i)$\cite{Delgado_Rossier_prb_2011}, where  $v_{\rm S}(i)$ and $v_{\rm T}(i)$ are  dimensionless factors that scale as
 the surface-adatom and tip-adatom hopping integrals.

 The quantum spin dynamics is described by means of a master equation for the diagonal elements of the density matrix\cite{Cohen_Grynberg_book_1998}, $P_M(V)$, described in the basis of eigenstates $|M\rangle$ of ${\cal H}_{\rm Spin}$.  
As we are interested in the inelastic spectroscopy of a dimer, we will explicitly write the inelastic contribution to the current
as\footnote{Readers interested in the details are referred to Ref.~\cite{Delgado_Rossier_prb_2010}, where general explicit expressions are provided.}
\begin{eqnarray}
I_{IN}=\frac{g_{S}}{G_0} \sum_{M,M',a}
 i_{-}(\Delta_{M,M'}+eV) \left| {\bf S}_{a,TS}^{MM'}\right|^2 P_M(V),
\label{iinelast}
\end{eqnarray}
with $\Delta_{M,M'}= E_M-E_{M'}$, bias voltage $V$, and the zero bias elastic conductance
$g_0\equiv (\pi^2/4)G_0 \rho_T \rho_S \left| T_0\right|^2\chi^2$, 
where  $G_0=2e^2/h$ is the quantum of conductance and $\chi=\sum_i v_T(i)v_S(i)$.
Here we have introduced the current associated to a single channel with energy $\Delta$ as
$i_{-}(\Delta+eV)= G_0/e\left({\cal G}(\Delta+eV) - {\cal G}(\Delta-eV)\right)$,
with  ${\cal G}(\omega)\equiv \omega/\left(1-e^{-\beta \omega}\right)$,
and the spin matrix elements
\begin{equation}
{\bf S}_{a,\eta\eta'}^{M,M'}\equiv \frac{1}{\chi}\sum_{i} v_{\eta}(i) v_{\eta'}(i) \langle M|{S}_a(i)|M'\rangle.
\label{transition-matrixW}
\end{equation}
Importantly, when both atoms are equally coupled to a given electrode, i.e., $v_\eta(i) =v_\eta$, the inelastic contribution is proportional to matrix elements of the total spin of the system.

\begin{figure}
\begin{center}
\includegraphics[angle=-90,width=0.49\linewidth]{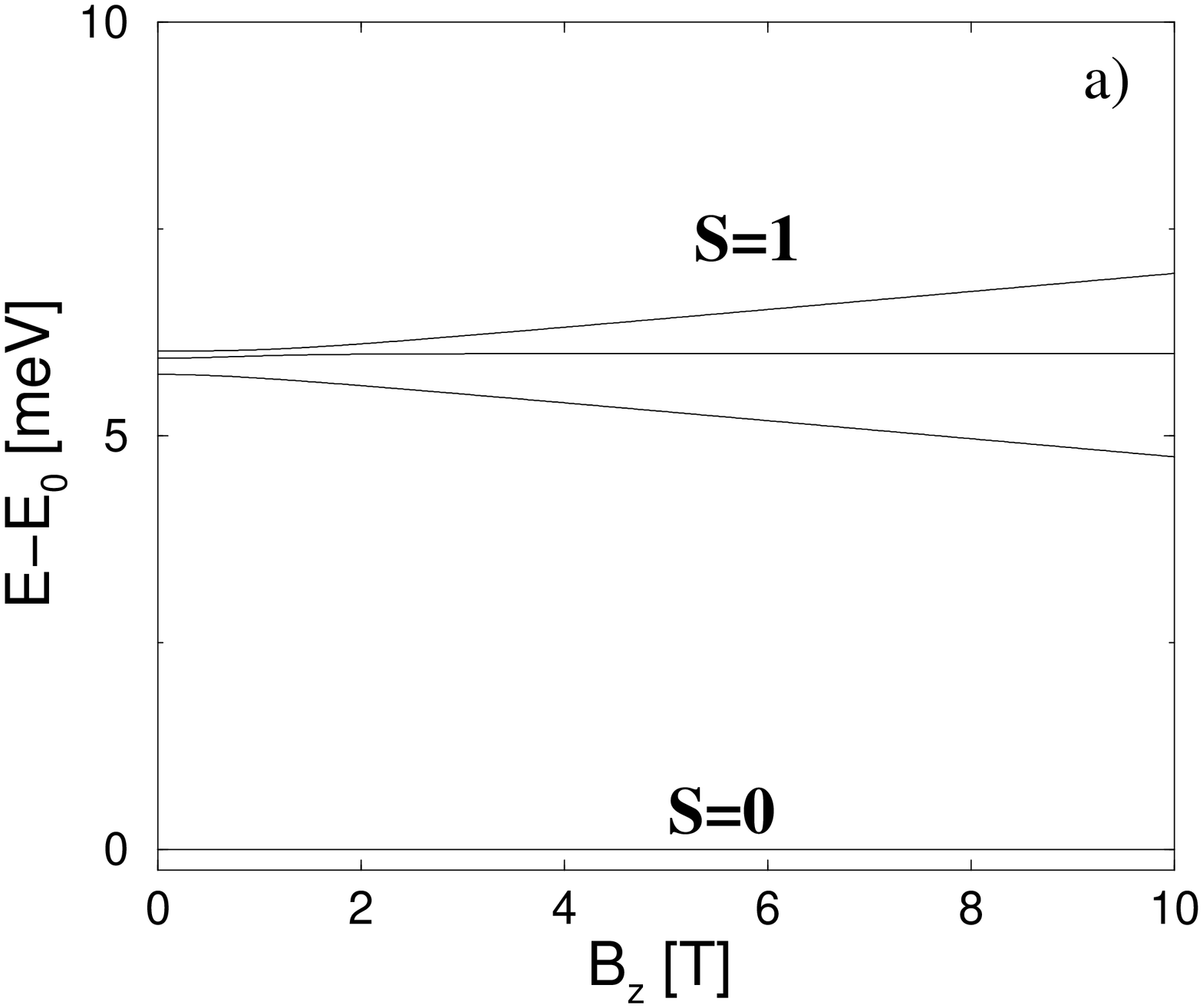}
\includegraphics[angle=-90,width=0.49\linewidth]{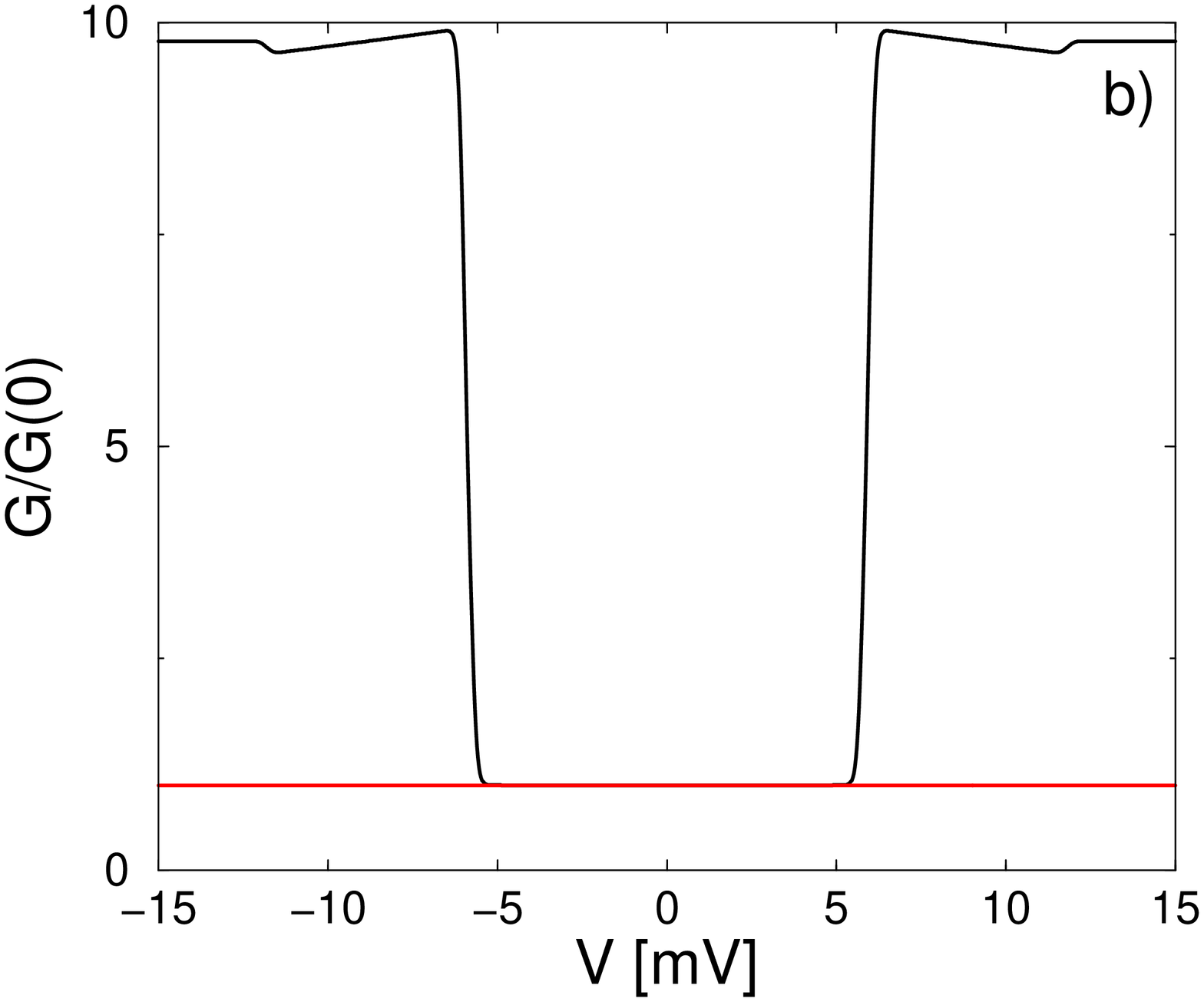}
\end{center}
\caption{ \label{fig1} (Color online) (a) Energy spectra corresponding to a Mn dimer
over a Cu$_2$N surface versus applied magnetic field.
 Spectrum is referred to the ground state energy. The magnetic
field is applied in the surface plane forming a $55º$ angle with the Cu-N direction\cite{Hirjibehedin_Lutz_Science_2006}.(b)
Corresponding $dI/dV$ curve for a non-polarized tip with two different tip couplings: $v_T(1)=0.1,\; v_T(2)=0$ (black line) and $v_T(i)=0.1$ (red line). Here $v_S(i)=1$, $T=0.6$K, and $T_0/{\cal T}=1$.}
\end{figure}
The Mn dimer has been studied experimentally \cite{Hirjibehedin_Lutz_Science_2006,Loth_Bergmann_natphys_2010} and theoretically\cite{Rossier_prl_2009,Delgado_Rossier_prb_2010}. In particular, Loth {\em et al.} observed a drastic modification of the line-shape with the amount of current, which have been explained in terms of non-equilibrium effects\cite{Delgado_Rossier_prb_2010}. 
The Mn-Mn exchange  interaction in this system is antiferromagnetic. The fitting\cite{Hirjibehedin_Lutz_Science_2006} of the experimental results to the 
Hamiltonian model, Eq. (\ref{hchain}), gives a $J_{1,2}\equiv J=5.9$, while $D=-0.039$ meV and $E=0.007$ meV, while $g=1.98$.

Fig.~\ref{fig1}(a) shows the lowest energy spectra of the Mn dimer. Since $J\gg |D|,E$,
the total spin $S$ is a good quantum number at zero order in $|D|/J$. The ground state is $S=0$, while the first excited state $S=1$ is located at an energy $\sim J$.  The $2S+1$ degeneracy of the $S>0$ multiplets is weakly lifted by the small anisotropy terms  $D$ and $E$. The allowed transitions induced by the exchange coupling (\ref{HTUN})   satisfy  $\Delta S=\pm 1$. At low current and for$v_T(1)\ne v_T(2)$, only the lowest energy excitation at $|eV|\approx J$ is observed, see Fig.~\ref{fig1}b), while for increasing current higher energy transitions appear\cite{Delgado_Rossier_prb_2010}.

The results above are apparently in very good agreement with the experimental data\cite{Loth_Bergmann_natphys_2010}. This is so if we assume that the exchange assisted tunneling is stronger through one of the atoms. However, the inelastic transitions  disappears
in the symmetric coupling case, $v_{{\rm T}}(1)=v_{{\rm T}}(2)$, see Fig.~\ref{fig1}b). 
This prediction of the model is in clear contrast with the experimental data\cite{Hirjibehedin_Lutz_Science_2006,Otte_PhD_2008}, which do not show a strong dependence of the inelastic current as the tip is moved along the Mn dimer axis. In fact, the topographic image of the Mn dimer shows that the STM is not capable of distinguishing both atoms: the dimer appears as a single and longer protuberance in the Cu$_2$N surface.
From the theoretical point of view, this cancellation arises from the fact that, in this particular case, the operator in the transition matrix element (\ref{transition-matrixW})  is the total spin of the dimer, and therefore, then the eigenstates of $S^2$ and $S_z$ are also eigenstates
of ${\cal V}$. 
 This problem appears not only in transport theories based on a exchange coupling between the localized spin and the transport electrons\cite{Appelbaum_pr_1967,Rossier_prl_2009,Fransson_nanolett_2009,Lorente_Gauyacq_prl_2009,Delgado_Palacios_prl_2010,Gauyacq_Novaes_prb_2010,Fransson_Eriksson_prb_2010,Zitko_Pruschke_njphys_2010}, but also in a two-sites Hubbard model\cite{Delgado_Rossier_prb_2011}. It is worth pointing out
that this problem is specific of the dimer. 
 In the case of a single Mn adatom, the observed spin transitions occur within states with the same $S=5/2$. For longer chains, the tip can not be coupled identically to all the atoms and the theory accounts for the data\cite{Rossier_prl_2009}.
 
\begin{figure}
\includegraphics[height=0.98\linewidth,width=0.7\linewidth,angle=-90]{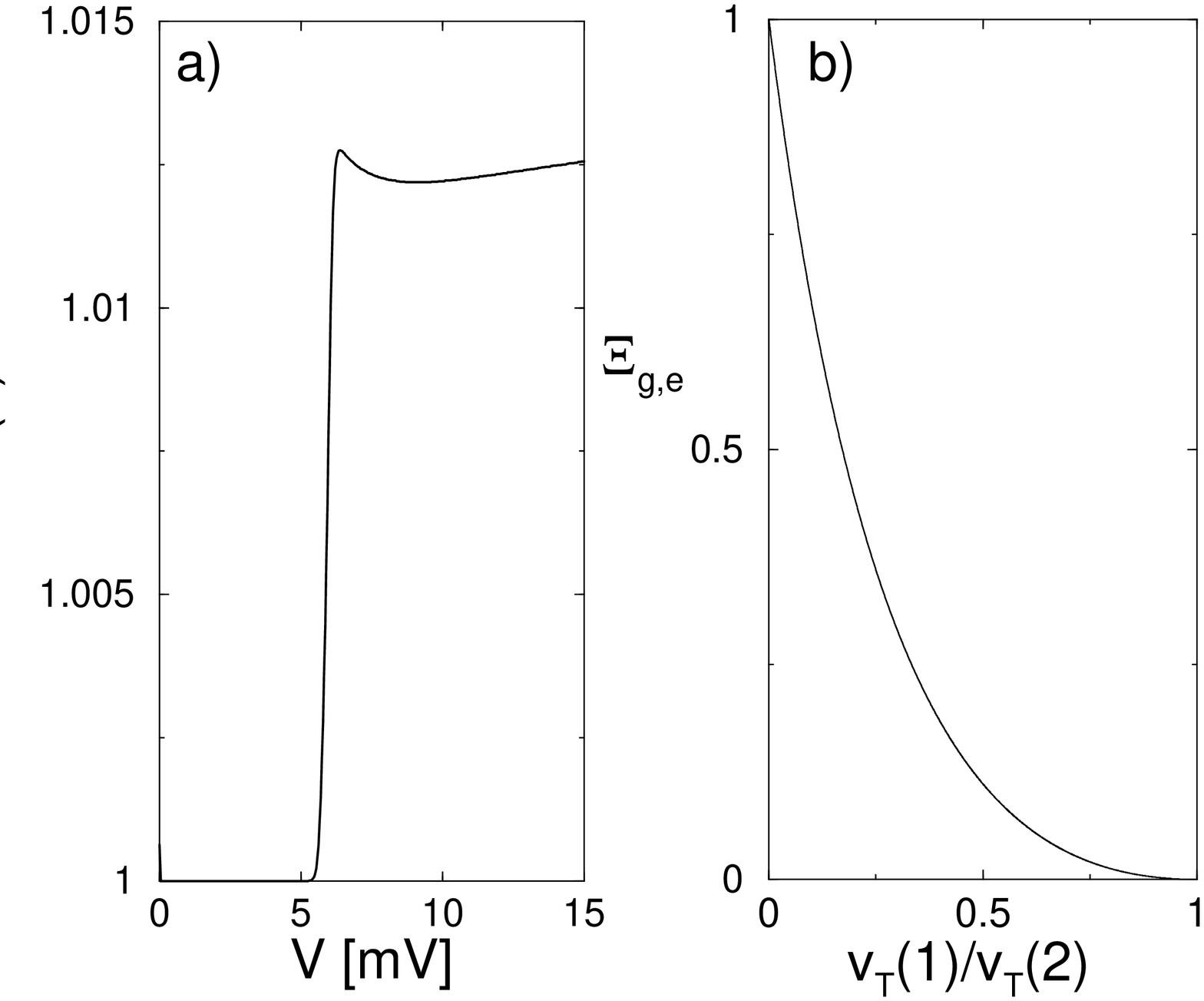}
\caption{a) $dI/dV$ curve for the symmetric coupling $v_T(i)=0.1$ and $v_S(i)=1$, including the hyperfine coupling to the nuclear spins.
 b)Transition amplitude  $\Xi_{S,T}$ versus the ratio $v_{{\rm T}}(1)/v_{{\rm T}}(2)$. 
 Other parameters as in Fig.~\ref{fig1}.
}
\label{fig2}
\end{figure}

In Fig. \ref{fig2}b) we plot the height of the inelastic step, given by
$\Xi_{S, T}=\sum_{a}\sum_{M'\subseteq S=1} \left|{\bf S}_{a,TS}^{S=0,M'} \right|^2$, normalized to its maximum value, as a function of the lateral position of the tip across the dimer axis, represented  by the ratio $v_{{\rm T}}(1)/v_{{\rm T}}(2)$. This strong dependence is not observed in the experiments\cite{Hirjibehedin_Lutz_Science_2006,Otte_PhD_2008}.

There are several spin interactions that break the spin rotational invariance and could, in principle, solve the problem. A promising candidate is the hyperfine coupling with the nuclear spin of the Mn. The only stable nuclear isotope of Mn is $^{55}$Mn   which has a nuclear spin $I=5/2$. Then, the electro-nuclear system of each Mn adatom has 36 states in total.  So, the corrected spin Hamiltonian should read as\cite{Delgado_Rossier_prl_2011}
\beqa
\label{hchainp}
{\cal H}^{^{55}{\rm Mn}}(i) = {\cal H}(i) +A{\bf S}(i).{\bf I}(i)\;.
\eeqa
The strength of the hyperfine coupling $A$ depends both on the nuclear magnetic moment and on the shape of the electronic cloud.  For the $^{55}$Mn isotope, $A$ typically ranges between 0.3 and 1$\mu$eV\cite{Walsh_Walter_PR_1965}.  Here we consider the largest value, $A\approx 1\mu eV$. In the case of Mn, the basic effect of the hyperfine coupling is to split each of the 6 electronic levels in ${\cal H}(i)$ into 6 nuclear branches. Although the hyperfine structure can not be resolved at the experimental temperature of $T\sim 0.6$ K\cite{Delgado_Rossier_prl_2011}, it has an important consequence: the total spin of the dimer is no longer conserved, opening new excitation and relaxation channels. Fig.\ref{fig2}a) shows the calculated $dI/dV$ curves including the hyperfine coupling for the symmetric coupling case. As observed, there is a finite step corresponding to the excitation of the first excited multiplet with  $S=1$. Nevertheless, the heigh of this step is much smaller than the observed experimentally.

\section{Summary}
To summarize, we have studied the effect of the hyperfine coupling in the IETS of a Mn dimer. We showed that this interaction is capable of breaking the conservation of the total spin and, therefore, induce a finite excitation, in the most likely experimental situation where both adatoms are equally coupled to the tip and substrate\cite{Hirjibehedin_Lutz_Science_2006,Otte_PhD_2008}. Unfortunately, the predicted heigh of the inelastic step is much weaker than in the experiments\cite{Hirjibehedin_Lutz_Science_2006,Loth_Bergmann_natphys_2010}.
This result suggests the presence 
 of additional terms in the tunneling Hamiltonian breaking the rotational invariance or the need to go beyond lowest order in perturbation theory. An alternative could be to do a microscopic calculation including all d-orbitals of both Mn adatoms, together with its interaction. This problem is a computational demanding calculation that will be let for a future work.

\begin{center}{\bf ACKNOWLEDGMENT }\end{center}
This work was supported by MEC-Spain (MAT07-67845,  FIS2010-21883-C02-01, 
Grants  JCI-2008-01885 and  CONSOLIDER CSD2007-00010) and Generalitat Valenciana (ACOMP/2010/070).




\end{document}